\documentstyle[prb,aps,epsf]{revtex}

\begin{document}
\twocolumn[\hsize\textwidth\columnwidth\hsize\csname @twocolumnfalse\endcsname

\title{Confined coherence and analytic properties of Green's
  functions}

\draft
\author {K.Sch\"onhammer}
\address{Institut f\"ur Theoretische Physik, Universit\"at
  G\"ottingen, Bunsenstr. 9, Germany}

\date{\today}
\maketitle

\begin{abstract}
A simple model of noninteracting electrons with a separable one-body
potential is used to discuss the possible pole structure of single particle
Green's functions for fermions  
on unphysical sheets in the complex frequency plane as a function of
the system parameters. The poles in the exact Green's function 
can cross the imaginary axis, in contrast to recent claims
that such a behaviour is unphysical. As the
Green's function of the model has the same functional form as
an {\it approximate} Green's function for coupled Luttinger
liquids no definite conclusions concerning the concept of "confined coherence",
can be drawn from the locations of the poles of this Green's function.

\end{abstract}
\vskip 2pc]
\vskip 0.1 truein
\narrowtext

Recently the concept of "confined coherence" was introduced in the
theoretical description of quasi-one dimensional conductors
\cite{CS}. Typically one considers two identical chains of strongly
interacting electrons which are prepared in their isolated chain
ground states but with one chain having $\delta N$ more electrons than
the other. A transverse hopping $t_{\perp}$ is switched on and the
behaviour of $<\delta N(t)>$ is studied. It is conjectured that  
the disappearance of interference oscillations in  $<\delta N(t)>$
represents a generic loss of coherence,
which implies that e.g. the single particle Green's function 
should not exhibit a pole on the real axis which disperses with the
perpendicular momentum. In this context Clarke and
Strong \cite{CS} examined the behaviour of the approximate Green's
function for coupled Luttinger liquids proposed by Wen \cite{W}. 
In their discussion these authors argue that if  
a pole crosses the imaginary axis on an unphysical sheet   
of the complex frequency plane this signals a transition between
qualitatively different phases of matter.
 In this short note we present 
simple model, where the {\em exact} Green's function in fact shows
such a behaviour without any qualitative change of the low
energy spectral properties at the crossing. This shows that it is
diffcult to draw definite conclusions from such a scenario.

We study the model of a half filled symmetric band of {\em
  noninteracting} spinless fermions perturbed by a {\em separable} one body
  potential

\begin{equation}\label{Hamiltonian}
H=\sum_{\bf k} \epsilon_{\bf k}c^{\dagger}_{\bf k}c_{\bf k}
+V_0a^{\dagger}a ,
\end{equation}
where the fermion operator $a$ is given by 
\begin{equation} \label{a}
a=\sum_{\bf k}\alpha_{\bf k}c_{\bf k},
\end{equation}
and the sum of $|\alpha_{\bf k}|^2$ is assumed to be unity.
Using the equation of motion method \cite{Z} the calculation of the single
particle Green's functions is straightforward.
For $\mbox{Im} z >0$ $(\mbox{Im} z <0)$ the retarded (advanced) 
Green's function
$G_{aa}(z)\equiv \langle\langle a;a^{\dagger}\rangle\rangle_z$  is given by
\begin{equation}\label{Gaa}
G_{aa}(z)=\frac{G^{(0)}_{aa}(z)}{1- V_0 G^{(0)}_{aa}(z)},
\end{equation}
with
\begin{equation}\label{Gaa0}
G^{(0)}_{aa}(z)=\int^B_{-B}\frac {\rho ^{(0)}_a(\epsilon)}{z-\epsilon}
d\epsilon,
\end{equation}
where the normalized spectral function $ \rho ^{(0)}_a(\epsilon) $
\begin{equation}\label{rho0}
 \rho ^{(0)}_a(\epsilon)=\sum_{\bf k}|\alpha_{\bf k}|^2
\delta(\epsilon -\epsilon_{\bf k})
\end{equation}
can be chosen arbitrarily.
We first assume this spectral weight to be largest in
the middle of band
\begin{equation}\label{powerlaw}
\rho ^{(0)}_a(\epsilon)=\frac {\gamma}{B^{2\gamma}}
\Theta(B^2-\epsilon^2)|\epsilon|^{2\gamma-1},
\end{equation}
where $\Theta(\cdot)$ is the step function and $B$ is half the bandwidth.  
We assume $\gamma$ to be in the range
 $0<2\gamma<1$. As $\rho^{(0)}_a(\epsilon)$ is symmetric, the
spectral
representation in Eq.(\ref{powerlaw})
can be written as 
\begin{equation}\label{spec}
G^{(0)}_{aa}(z)=2z\frac {\gamma}{B^{2\gamma}}\int ^B_{0}
\frac{\epsilon^{2\gamma-1}}{z^2-\epsilon^2}d\epsilon
=z\frac {\gamma}{B^{2\gamma}} \int^{B^2}_0
\frac {u^{\gamma-1}}{z^2-u}
du.
\end{equation}
In the following we discuss {\it retarded} Green's functions which are
analytic in the {\it upper} half plane. On the positive imaginary axis
$G^{(0)}_{aa}$ is given by 
\begin{equation}\label{iy}
G^{(0)}_{aa}(iy)=-iy\frac {\gamma}{B^{2\gamma}}
\int^{B^2}_0 \frac {u^{\gamma-1}}{y^2+u}du.
\end{equation}
Assuming  $0<y \ll B$ and using our assumption about $ \gamma$, 
the upper limit of
the integration can be replaced by $\infty$. Then $ G^{(0)}_{aa}(iy) $
can be calculated analytically
\begin{equation}\label{iy2}
 G^{(0)}_{aa}(iy)=\frac{1}{iy}\left (\frac{-(iy)^2}{B^2}\right )^{\gamma}
\frac{\gamma\pi}
{\sin\gamma\pi }.
\end{equation}
For arbitrary $\mbox{Im} z>0$
 we just have to replace $iy$ by $z$ 
\begin{equation}\label{RGF}
G^{(0)}_{aa}(z)=\frac{1}{z}\left (\frac{-z^2}{B^2}\right )^{\gamma}
\frac{\gamma\pi}{\sin\gamma\pi }.
\end{equation}
For $\gamma$ irrational this function has infinitely many Riemann
sheets  and one has to identify the "physical sheet" defined by
Eq. (\ref{Gaa0}). If we write $z=|z|e^{i\phi}$ we have
\begin{equation}\label{Phase}
(-z^2)^{\gamma}=|z|^{2\gamma}e^{i\left [
\frac {\pi}{2}(1+2n)+\phi\right ] 2\gamma},
\end{equation}
where $n \in \mathbf Z$ has to be chosen properly. We use the fact that the
phase of $G^{(0)}_{aa}$
has to be $-\pi/2$ on the positive imaginary axis as seen from
Eq. (\ref{iy}).
This requires $n=-1$, i.e.
\begin{equation}\label{Riemann}
G^{(0)}_{aa}(|z|,\phi)=\frac {1}{B}\left ( \frac{|z|}{B}\right
)^{2\gamma-1}
\frac {\gamma\pi}{\sin\gamma\pi}e^{i\Phi_R(\phi)},
\end{equation}
with $ \Phi_R(\phi)=(\phi-\pi/2)2\gamma-\phi$ . It is easy to check
that $-\mbox{Im} G^{0}_{aa}/\pi$ is different from zero for $0<\phi<\pi$
and yields the spectral function in Eq. (\ref{powerlaw}) when $z$
approaches the real axis from above. If $\phi$ in Eq. (\ref{Riemann})
is allowed to vary from minus to plus infinity one obtains the
function
on the complete Riemann surface.

The fact that  $-\mbox{Im} G^{0}_{aa}/\pi$ is different from zero for
$0<\phi<\pi$
implies that the
 full Green's function $G_{aa}(z)$ in Eq. (\ref{Gaa}) is analytic
for $0<\phi<\pi$ as every exact retarded Green's function. It has poles
on the unphysical sheets corresponding to $\phi$ values outside this range.
In the following we discuss a {\it repulsive} interaction $V_0>0$.
Then a necessary condition for $G_{aa}(z)$
to have a pole is $ \Phi_R(\phi_{(m)})=2\pi m $, where $m$ is integer,
i.e. $\phi_{(m)}=-(\pi\gamma+2\pi m)/(1-2\gamma)$.
The absolute value of $z$ at the poles is independent of $m$ and
given by
\begin{equation}\label{zB}
|z_{(m)}|=B\left (\frac{V_0}{B} \frac {\gamma\pi}{\sin\gamma\pi} \right )
^{\frac{1}{1-2\gamma}},
\end{equation}
\begin{figure}[hbt]
\hspace{0.0cm}
\epsfxsize8.0cm
\epsfbox{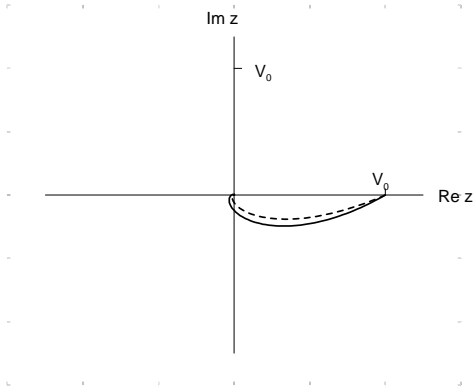}
\vspace{0.0cm}
\caption{Projection of the $m=0$ pole location 
 on the complex frequency plane as a function of $\gamma$
for different values of $V_0/B$ (dottet line: $V_0/B=0.1$, dashed
line: $V_0/B=0.05$). For infinitesimal $\gamma$ the pole is at
$V_0-i0$, and   for $\gamma \to 1/2$
approaches the origin circling it on the different sheets
of the Riemann surface.}
\end{figure}
which reduces to $V_0$ for $\gamma \to 0$.
For infinitesimal $\gamma$ the pole for $m=0$ is in the forth quadrant
just below the real axis. As shown in Fig. 1 for two different values
of $V_0/B$
 it moves away from the
real axis when $\gamma$ is increased, crosses the negative imaginary
axis for $\gamma=1/4$ and circles the origin of the complex $z-$plane
infinitely many times
when approaching $\gamma=1/2$. The latter fact cannot really be seen
in
the figure as $|z_{(0)}|$ goes to zero very quickly near $\gamma=1/2$.   

\begin{figure}[hbt]
\hspace{0.0cm}
\epsfxsize8.0cm
\epsfbox{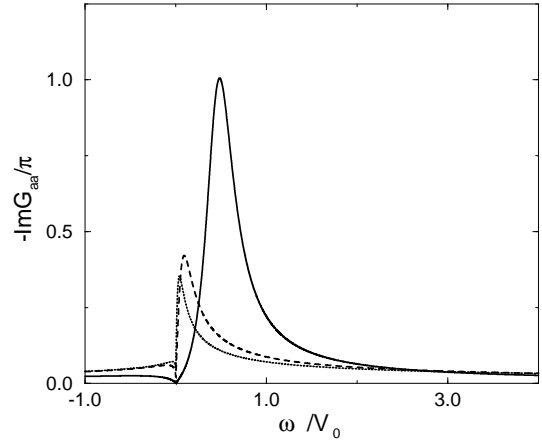}
\vspace{0.0cm}
\caption{Spectral function of the exact Green's function $G_{aa}$
as a function of $\omega/V_0$
for $V_0/B=0.05$ and different values of $\gamma$ (full line:
$\gamma=0.1$, dashed line: $\gamma=0.23$, dotted line: $\gamma=0.27$).
No qualitative change occurs at $\gamma=1/4$ , where the 
pole on the unphysical sheet crosses the negative imaginary axis.  }
\end{figure}
The spectral function $-\mbox{Im} G_{aa}(\omega+i0)/\pi$ is shown in Fig.2
for different values of $\gamma$ with $V_0/B$ fixed.
Changing this ratio only changes the scales of the functions.
For $\gamma \ll 1$ a well defined peak occurs with the position and
width determined by the position of the pole on the unphysical sheet.
With increasing $\gamma$ the weight for negative $\omega$ and for
$\omega \gg V_0$ increases. {\it No} qualitative change occurs when
the pole on the unphysical sheet crosses the negative imaginary axis
for $\gamma=1/4$. To show this we present results for $\gamma=0.23$
and $\gamma=0.27$. Also for $1/4<\gamma<1/2$ the larger weight remains
in the positive frequency range. The pole on the unphysical sheet 
determines the spectral function in an essential way {\it only} if it is 
close to the physical sheet and well
separated from all other non-analyticities like the branch point
in the origin.

As a second example we discuss an {\it asymmetric} spectral density
\begin{equation}\label{rho2}
\rho^{(0)}_a(\epsilon)=\frac{\gamma}{(B^2-\epsilon^2_0)^{\gamma}}
(\epsilon^2-\epsilon^2_0)^{\gamma}/|\epsilon-\epsilon_0|,
\end{equation}
for $\epsilon^2_0<\epsilon^2<B^2$ and zero elsewhere,
where $0<\epsilon_0\ll B$. It has a small gap from $-\epsilon_0$ to
$\epsilon_0 $ in the center of the band and a power law divergence 
when $\epsilon $ approaches $\epsilon_0$ from above.
If $\epsilon^2_0$ is neglected compared to $B^2$ 
the corresponding Green's function $G^{(0)}_{aa}$ 
for $|z| \ll B$ is obtained by
replacing $1/z$ in Eq. (\ref{RGF}) by $1/(z-\epsilon_0)$ and
$z^2$ in the parenthesis by $z^2-\epsilon^2_0$. 
The  correct phase can be determined similarly to the $\epsilon_0=0$
case . Figure 3 shows corresponding spectral functions 
 $-\mbox{Im} G_{aa}(\omega+i0)/\pi $ for a repulsive one-body
potential for two different values of $\gamma$. For the smaller value
$\gamma=0.3$ the real part of the pole on the unphysical sheet
is larger than $\epsilon_0$ while for $\gamma=0.36$ it is smaller
than $\epsilon_0$. This leads to {\it no} qualitative difference in the
spectral functions.

\begin{figure}[hbt]
\hspace{0.0cm}
\epsfxsize8.0cm
\epsfbox{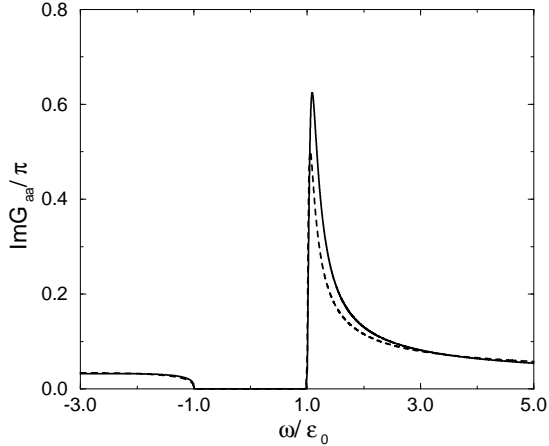}
\vspace{0.0cm}
\caption{Spectral function of the exact Green's function $G_{aa}$
for the asymmetric spectral density in Eq.(\ref{rho2}) and a
repulsive interaction. The parameters chosen are $V_0/\epsilon_0=0.5 $ and 
$V_0/B=0.05$. The full (dotted) curve corresponds to $\gamma=0.3$
($\gamma=0.36$).} 
\end{figure}

For attractive one-body potentials the spectral functions
$-\mbox{Im}G_{aa}
/\pi$ have an additional  
delta peak in the gap which results from a pole on the real
axis on the physical sheet. Its weight vanishes for $|V_0|/\epsilon_0
\to 0$ proportional to $(|V_0|/\epsilon_0)^{\gamma/(1-\gamma)} $.

The low energy form Eq.(\ref{RGF}) of the 
Green's function $G^{(0)}_{aa}$ of our simple model of 
noninteracting electrons has the same form as the {\it low
energy} Green's function $g(k_F,z)$ of interacting fermions in one
dimension \cite{W}, when $2\gamma$ is identified with the anomalous
dimension $\alpha$.
 If one puts $\epsilon_0=v_c(k_{\parallel}-k_F)$ in Eq. (\ref{rho2})
this corresponds to the 
case $k_{\parallel}\neq k_F$ for spinless fermions \cite{W}. 
 If a perpendicular hopping $t_{\perp}$ between a
lattice of chains 
is treated in leading order perturbation theory in $t_{\perp}$ the
resulting approximate Green's function \cite{W} of the coupled system for
${\bf k}=(k_F,{\bf k}_{\perp})$ has the form of $G_{aa}$ in Eq.
(\ref{Gaa}) when $V_0$ is replaced by $t_{\perp}({\bf k}_{\perp})$.
This approximate Green's function \cite{W} for coupled Luttinger
liquids implies Fermi liquid behaviour for $\alpha<1$ and 
``confinement'' for $\alpha>1$, in agreement with renormalization group
arguments. It should be pointed out that for $\alpha<1$ 
this 
Green's functions shows various deviations from usual Fermi liquid
behaviour.
 
Clarke and Strong \cite{CS} present evidence that the 
interference oscillations $<\delta N(t)>$ disappear for a value of the
anomalous dimension $\alpha=1/2$ (i.e. $\gamma=1/4$)
or smaller. They expect to see this ``transition'' also in the
behaviour of the one-particle Green's function. As Wen's Green's
function for $k_{\parallel}=k_F$
shows the crossing of the negative imaginary axis of
the pole on the unphysical sheet
at $\gamma=1/4$
they take this as evidence for their conjecture. Our simple model 
shows that there is {\it no} general physical principle to forbid 
such a crossing
and therefore at least in Wen's approximation
the single particle Green's function shows {\it no } sign of a sudden 
loss of
coherence if it in fact occurs for $\alpha<1$.  

The author would like to thank P.Kopietz and V.Meden for discussions
and useful suggestions on the manuscript.

\end{document}